\documentclass[11pt]{article}
\usepackage{amssymb,amsmath,cite}
\catcode`\@=11

\@addtoreset{equation}{section}
\def\theequation{\arabic{section}.\arabic{equation}}
     
\catcode`\@=11
\def\thesection{\arabic{section}}

\def\appendix{\setcounter{section}{0}
        \def\thesection{Appendix}
        \def\theequation{\Alph{section}.\arabic{equation}}}
\def\section{\@startsection{section}{1}{\z@}{3.5ex plus 1ex minus
   .2ex}{2.3ex plus .2ex}{\large\bf}}
   
\long\def\@makefntext#1{\parindent 0cm\noindent
\hbox to 1em{\hss$^{\@thefnmark}$}#1}

\newcommand{\captionfonts}{\small}
\makeatletter  
\long\def\@makecaption#1#2{%
  \vskip\abovecaptionskip
  \sbox\@tempboxa{{\captionfonts #1: #2}}%
  \ifdim \wd\@tempboxa >\hsize
    {\captionfonts #1: #2\par}
  \else
    \hbox to\hsize{\hfil\box\@tempboxa\hfil}%
  \fi
  \vskip\belowcaptionskip}
\makeatother   

\begin{document}
\begin{titlepage}
\vspace{.5in}
\begin{flushright}
September 2008\\
\end{flushright}
\vspace{.5in}
\begin{center}
{\Large\bf
The Constraint Algebra\\[1ex]
of Topologically Massive AdS Gravity} 

\vspace*{.4in}
{S.~C{\sc arlip}\footnote{\it email: carlip@physics.ucdavis.edu}\\
       {\small\it Department of Physics}\\
       {\small\it University of California}\\
       {\small\it Davis, CA 95616}\\{\small\it USA}}
\end{center}

\vspace{.5in}
\begin{center}
{\large\bf Abstract}
\end{center}
\begin{center}
\begin{minipage}{4.5in}
{\small
Three-dimensional topologically massive AdS gravity has a complicated 
constraint algebra, making it difficult to count nonperturbative degrees 
of freedom.  I show that a new choice of variables greatly simplifies this 
algebra, and confirm that the theory contains a single propagating mode 
for all values of the mass parameter and the cosmological constant.  As an 
added benefit, I rederive the central charges and conformal weights 
of the boundary conformal field theory from an explicit analysis of the 
asymptotic algebra of constraints.
}
\end{minipage}
\end{center}
\end{titlepage}
\addtocounter{footnote}{-1}

\section{Introduction} \label{intro}

Topologically massive gravity \cite{DJT}---(2+1)-dimensional Einstein 
gravity supplemented with a Chern-Simons term for the spin connection%
---provides a fascinating playground for exploring higher-derivative 
gravity.   In contrast to the topological character of ordinary Einstein 
gravity in three dimensions, topologically massive gravity has a local 
degree of freedom, a parity-violating massive spin two graviton that can 
be described by a single indexless ``scalar'' field.  With the addition of a 
negative cosmological constant $\Lambda=-1/\ell^2$, surprising new 
features emerge \cite{CDWW}: for instance, the components of curvature 
perturbations propagate with different, chirality-dependent masses.

Topologically massive AdS gravity may also provide a useful model
in which to explore the AdS/CFT correspondence.  The conformal
boundary of a three-dimensional asymptotically anti-de Sitter spacetime 
is a flat two-dimensional cylinder, and the asymptotic symmetries are
described by a pair of Virasoro algebras \cite{BH}.  The resulting
two-dimensional conformal symmetry can be very powerful.  In pure Einstein
gravity, for example, although the boundary conformal field theory is 
not known \cite{Carlip,Witten}, the classical central charges and
conformal weights are sufficient to determine the BTZ black hole
entropy \cite{Strominger,BSS} and even the spectrum of Hawking 
radiation \cite{Emparan}.  In topologically massive gravity, it was
shown several years ago that the central charges of the ``left'' and 
``right''  Virasoro algebras split \cite{KL,Solodukhin,Hotta}:
\begin{equation}
c_\pm = \frac{3\ell}{2G}\left( 1 \pm \frac{1}{\mu\ell} \right),
\label{intro1}
\end{equation}
where the mass parameter $\mu$ is fixed by the Chern-Simons coupling.
The classical contributions to conformal weights are also shifted, 
leading to interesting modifications of black hole thermodynamics 
\cite{KL,Solodukhin,Park}.  Moreover, at $\mu\ell=\pm1$ the boundary
theory becomes chiral.  Since the sum over topologies in three-dimensional
gravity may require a chiral splitting \cite{Maloney}, such a theory could 
be of considerable interest.

Unfortunately, this model appears to have a fundamental sickness.  With 
the usual sign for the gravitational constant, the massive excitations of 
topologically massive gravity carry negative energy \cite{DJT}.  In the 
absence of a cosmological constant, one can simply flip the sign of $G$,
but if $\Lambda<0$, this will give a negative mass to the BTZ black hole
\cite{Moussa}.  The existence of a stable ground state is thus in doubt.
The possibility of a supersymmetric extension of the theory \cite{Deser}
suggests the existence of a stable superselection sector, but this sector
appears to exclude black holes.

Recently, Li, Song, and Strominger proposed a possible cure \cite{LSS}.
At the chiral point, a family of eigenstates of the Virasoro generator $L_0$
representing massive excitations disappears, and Li et al.\ suggested 
that the massive gravitons might no longer be present.  Unfortunately, a
different family of finite-energy eigenstates of $L_0$ has been found 
\cite{GJ}, which violate the standard Fefferman-Graham asymptotic 
conditions \cite{FG} but are still asymptotically anti-de Sitter; and worse, 
a complete set of finite-energy asymptotically AdS wave packets also 
exists, even at the chiral coupling \cite{CDWW}.  

A loophole remains, however.  The computations of \cite{LSS,GJ,CDWW}%
---and, indeed, those of virtually every paper discussing this model---are based on 
classical perturbation theory, expanding the metric in small fluctuations 
around AdS and keeping only lowest order terms.   The full field equations, 
on the other hand, are highly nonlinear, and it is conceivable that new 
features could emerge nonperturbatively.

A general nonperturbative solution of the field equations of topologically
massive gravity seems distant, but we can learn a great deal by
analyzing the constraints.  For the case of a vanishing cosmological
constant, such an analysis was first performed by Deser and Xiang \cite{DX}, 
and further amplified by Buchbinder et al.\ \cite{Buchbinder}.  The
formalism is very complicated, in part because of the presence of third
derivatives and second class constraints, but the results ultimately confirm 
the existence of a single propagating degree of freedom.  For the asymptotically 
anti-de Sitter case, the literature is currently inconsistent: Park \cite{Parkb}
appears to find more than one degree of freedom (one ``for each internal
index''), while Grumiller et al.\ \cite{GJJ} find one configuration space 
degree of freedom, but consider only the chiral coupling.

In this paper, I will show that a new choice of variables greatly simplifies the
constraint analysis, allowing an elegant expression of the constraint algebra
and a simple counting of degrees of freedom.  I confirm the existence of a
single propagating degree of freedom at all values of the couplings, and  
rederive the central charges (\ref{intro1}) from an explicit computation of
the algebra of asymptotic symmetries.

\section{Topologically massive AdS gravity}

In the first order formalism of (2+1)-dimensional gravity, the fundamental
variables are the triad $e^a = e^a{}_\mu dx^\mu$ and the spin connection 
$\omega_a = \frac{1}{2}\epsilon_{abc}\omega^{bc}{}_\mu dx^\mu$.
The Einstein-Hilbert action takes the form\footnote{I choose units 
$16\pi G=1$, a metric of signature $(-++)$, and a cosmological constant
$\Lambda=-1/\ell^2$, and set $\epsilon_{012}=1$ (so $\epsilon^{012}=-1$).}
\begin{equation}
I_{EH}[e,\omega] = 2 \int \left[ 
    e^a\wedge\left( d\omega_a 
    + \frac{1}{2}\epsilon_{abc}\omega^b\wedge\omega^c\right)
    + \frac{1}{6}\frac{1}{\ell^2}\epsilon_{abc}e^a\wedge e^b\wedge e^c\right] ,
\label{b1}
\end{equation}
where $e$ and $\omega$ can be treated as independent variables.  The variation 
of $\omega$ yields the torsion constraint
\begin{equation}
T_a = D_\omega e_a = de_a + \epsilon_{abc}\omega^b\wedge e^c  = 0 ,
\label{b2}
\end{equation}
while the variation of $e$ gives the Einstein field equations.   

An additional Chern-Simons term can be written for the spin connection,
\begin{equation}
I_{CS}[\omega] =  \int\left[
    \omega^a\wedge\left( d\omega_a 
    + \frac{1}{3}\epsilon_{abc}\omega^b\wedge\omega^c\right)\right] .
\label{b3}
\end{equation}
If $e$ and $\omega$ are varied independently, the sum of the Einstein-Hilbert
and Chern-Simons actions gives a model whose solutions are identical to those of
ordinary Einstein gravity \cite{Wittenb}, although with a different symplectic 
structure that may have implications for the quantum theory \cite{Meusburger}.  
If the torsion constraint (\ref{b2}) is imposed, however, one obtains a higher-derivative
theory, topologically massive gravity, with  
\begin{equation}
I_{TMG}[e] = I_{EH}[e,\omega(e)] + \frac{1}{\mu}I_{CS}[\omega(e)] .
\label{b4}
\end{equation}

To simplify this action, let us define a new connection
\begin{equation}
A^a = \omega^a + \mu e^a  ,
\label{b5}
\end{equation}
whose Chern-Simons action is
$$
\frac{1}{\mu} I_{CS}[A] = \frac{1}{\mu} I_{CS}[\omega] + I_{EH}
    + \int \left[ \mu e^a\wedge T_a + \frac{1}{3}\left(\mu^2 - \frac{1}{\ell^2}\right)
    \epsilon_{abc}e^a\wedge e^b\wedge e^c\right] .
$$
Rather than explicitly writing $\omega$ as a function of $e$, we can
impose the torsion constraint with a Lagrange multiplier, as suggested in
\cite{DX,Carlipb}.  Then
\begin{equation}
I_{TMG} = \frac{1}{\mu} I_{CS}[A] 
    + \int \left[ \beta^a\left( D_A e_a - \mu \epsilon_{abc}e^b\wedge e^c\right)
    - \alpha\epsilon_{abc}e^a\wedge e^b\wedge e^c\right] ,
\label{b6}
\end{equation}
where  $\beta^a = \beta^a{}_\mu dx^\mu$ is a Lagrange multiplier for the
torsion constraint and 
$$
\alpha = \frac{1}{3}\left(\mu^2 - \frac{1}{\ell^2}\right) .
$$ 
Note that the chiral coupling occurs at $\alpha=0$; this is the only place in which the 
relationship of $\mu$ and $\ell$ appears in this formulation.

The classical equations of motion may be obtained by varying $A$, $\beta$, and $e$:
\begin{align}
\delta A: \qquad& F_a + \frac{\mu}{2}\epsilon_{abc}\beta^b\wedge e^c = 0 
    \qquad \hbox{with}\  F_a = dA_a + \frac{1}{2}\epsilon_{abc}A^b\wedge A^c 
    \nonumber\\
\delta\beta: \qquad& T_a = D_A e_a - \mu \epsilon_{abc}e^b\wedge e^c = 0 
     \nonumber\\
\delta e: \qquad&B_a 
    = D_A\beta_a - 2\mu\epsilon_{abc}\beta^be^c - 3\alpha\epsilon_{abc}e^be^c = 0 .
    \label{b7} 
\end{align}
I show in the Appendix that these are equivalent to the standard field equations for
topologically massive AdS gravity, and that they determine the Lagrange multiplier
$\beta$ to be 
\begin{align}
&\beta^a{}_\mu =
  -\frac{2}{\mu}\left(R^a{}_\mu + \frac{2}{\ell^2}e^a{}_\mu 
  + \frac{3\alpha}{2} e^a{}_\mu\right) 
   \nonumber\\
&\beta = \beta^a{}_\mu e_a{}^\mu = -\frac{9\alpha}{\mu}  . \label{b9}
\end{align}

\section{Poisson brackets and constraints}\label{constraints}

From the action (\ref{b6}), we can now read off the terms involving time derivatives:
\begin{equation}
I_{TMG} = \int d^3x \epsilon^{ij}\left[ -\frac{1}{\mu}A^a{}_i\partial_t A_{aj} 
    - \beta^a{}_i\partial_t e_{aj}\right] + \dots
\label{c1}
\end{equation}
The canonical Poisson brackets are thus
\begin{align}
&\left\{ A^a{}_i, A^b{}_j\right\} = \frac{\mu}{2}\eta^{ab}\epsilon_{ij} \nonumber\\
&\left\{ e^a{}_i, \beta^b{}_j\right\} = \eta^{ab}\epsilon_{ij} .\label{c2}
\end{align}
The factor of $1/2$ in the first bracket can be obtained from Dirac brackets, or more
simply by recognizing that $A_x$ and $A_y$ are conjugate and integrating by parts.  This
diagonalization of the Poisson brackets is one of the main simplifications coming from our 
choice of variables.

The time components $A^a{}_t$, $e^a{}_t$, and $\beta^a{}_t$ appear in the action 
without time derivatives, and can be considered Lagrange multipliers.  The corresponding 
constraints are just the relevant components of the classical equations of motion:
\begin{align}
&J_a = -\frac{2}{\mu}\epsilon^{ij}\left(
    F_{aij} + \frac{\mu}{2}\epsilon_{abc}\beta^b{}_i e^c{}_j\right)  
   \nonumber\\
&T_a = -\epsilon^{ij}\left(D_i e_{aj} - \mu \epsilon_{abc}e^b{}_i e^c{}_j\right) 
    \nonumber \\
&B_a 
    = -\epsilon^{ij}\left(D_i\beta_{aj} - 2\mu\epsilon_{abc}\beta^b{}_ie^c{}_j 
    - 3\alpha\epsilon_{abc}e^b{}_ie^c{}_j\right) . 
\label{c3} 
\end{align}
It is  convenient to ``smear'' the constraints, integrating them against vectors.  More
precisely, let us define
\begin{align}
&J[\xi] = \int_\Sigma d^2x\, \xi^aJ_a + Q_J[\xi]  \nonumber\\
&T[\xi] = \int_\Sigma d^2x\, \xi^aT_a + Q_T[\xi]  \nonumber\\
&B[\xi]  =  \int_\Sigma d^2x\, \xi^aB_a + Q_B[\xi] ,
\label{c4}
\end{align}
where $Q_J$, $Q_T$, and $Q_B$ are the boundary terms needed to make the constraints
differentiable \cite{BH}.   For now, we shall assume that the parameters $\xi^a$ fall 
off rapidly enough at infinity that we can freely integrate by parts and  ignore
boundary terms; these will be restored below in section \ref{BT}.  The Poisson brackets 
of these generators with the canonical variables $\{A,e,\beta\}$ are now easy to compute,
and are displayed explicitly in the Appendix.
 
The constraints $\{J,T,B\}$ should generate the full group of symmetries of the
action (\ref{b6}), including diffeomorphism invariance.  For (2+1)-dimensional 
Einstein gravity, it is known that an appropriate combination of the ``gauge''
constraints with field-dependent parameters does, indeed, generate
diffeomorphisms on shell \cite{Wittenb,Banados0,Banados}.  The same is true here.  Let 
$\xi^\mu$ be an arbitrary three-vector, and define
\begin{equation}
 \xi^a = e^a{}_\mu\xi^\mu, \qquad {\hat\xi}^a = \beta^a{}_\mu\xi^\mu, \qquad 
    {\tilde\xi}^a = A^a{}_\mu\xi^\mu .
\label{c6}
\end{equation}
Define a new combination $H$ of the constraints by
\begin{equation}
H[\xi] = B[\xi] + T[{\hat\xi}] + J[{\tilde\xi}] .
\label{c7}
\end{equation}
A simple computation then shows that
\begin{equation}
\{H[\xi],X\} = -{\cal L}_\xi X + \hbox{\it terms proportional to the equations of motion,}
\label{c8}
\end{equation}
where $X$ is any of $\{A,e,\beta\}$ and $\cal L$ denotes the Lie derivative.  

\section{The algebra of constraints}\label{algebra}

We next turn to the algebra of the constraints.  We may first check that $J[\xi]$ generates 
local Lorentz transformations: a straightforward calculation shows that for any constraint 
$C[\eta]$,
\begin{equation}
\left\{ J[\xi],C[\eta]\right\} 
    = -C[\xi\times\eta] \quad \hbox{with $(\xi\times\eta)^a=\epsilon^{abc}\xi_b\eta_c$} .
\label{d1}
\end{equation}
The $J[\xi]$ are thus first class constraints.

The remaining constraints are more complicated.  We find that
\begin{align}
\left\{ T[\xi],T[\eta]\right\} &=
    -\frac{\mu}{2}\int d^2x\,\xi^a\eta^b\left( \epsilon^{ij}e_{ai}e_{bj}\right) \nonumber\\
\left\{ B[\xi],T[\eta]\right\} &= -\frac{\mu}{2}J[\xi\times\eta] + 2\mu T[\xi\times\eta]
    + \frac{\mu}{2}\int d^2x\,\xi^a\eta^b\left(  \epsilon^{ij}\beta_{ai}e_{bj}  
    -  \eta_{ab}\epsilon^{ij}\beta^c{}_ie_{cj}\right)   \nonumber\\
\left\{ B[\xi],B[\eta]\right\} &= 2\mu B[\xi\times\eta] + 6\alpha T[\xi\times\eta]
   -  \frac{\mu}{2}\int d^2x\,\xi^a\eta^b\left(\epsilon^{ij}\beta_{ai}\beta_{bj}\right) .
   \label{d2}
\end{align}
The appearance on the right-hand side of terms that are not proportional to the 
constraints can mean two things: either there are secondary constraints, or some of our
constraints are second class \cite{HT}.  To distinguish the two possibilities, note first that
$\epsilon^{ij}e_{ai}e_{bj}$ is zero only if the triad is noninvertible, certainly
not a restriction we wish to impose.  Similarly, $\epsilon^{ij}\beta_{ai}\beta_{bj}$
and $\epsilon^{ij}e_{ai}\beta_{bj}$ vanishes only if some components of $\beta_{ai}$ 
are linearly dependent.  From (\ref{b9}), this is not generically true, holding only for 
``pure Einstein gravity'' solutions.  

On the other hand, the quantity
$$
\Delta = \epsilon^{ij}\beta^c{}_i e_{cj} = \epsilon^{ij}\beta_{ij}
$$
always vanishes classically, by virtue of the symmetry of $\beta_{\mu\nu}$.   We can 
therefore treat $\Delta$ as a secondary constraint,\footnote{Alternatively, we could  
treat the Hamiltonian as a second class constraint; as discussed in \cite{BGH}, the rank 
of the brackets of the primary constraints would then no longer be constant, and the Dirac 
brackets would become singular in some regions of phase space.}  and obtain the further 
commutators
\begin{align}
&\left\{ T[\xi],\Delta\right\} = -\epsilon^{ij}D_i\left(\xi^ae_{aj}\right) 
     - \mu\epsilon^{abc}\left(\epsilon^{ij}e_{ai}e_{bj}\right)\xi_c  - \xi^aT_a  
    \nonumber\\
&\left\{ B[\xi],\Delta\right\} = \epsilon^{ij}D_i\left(\xi^a\beta_{aj}\right) 
     - 2\mu\epsilon^{abc}\left(\epsilon^{ij}\beta_{ai}e_{bj}\right)\xi_c
     - 9\alpha\epsilon^{abc}\left(\epsilon^{ij}e_{ai}e_{bj}\right)\xi_c + \xi^aB_a 
    \nonumber\\
&\left\{ \Delta(x),\Delta(x')\right\} = 0 ,\label{d3}
\end{align}
along with the relation $\{J[\xi],\Delta\}=0$ that we would expect from the role of
$J$ as a generator of local Lorentz transformations.

Our task is now to diagonalize the Poisson brackets (\ref{d2})--(\ref{d3}), to determine
the first and second class constraints.  To do so, let us define ${\hat\xi}^a$ 
to be such that 
$$
e_{ai}{\hat\xi}^a = \beta_{ai}\xi^a ,
$$  
and let
\begin{equation}
{\hat B}[\xi] = B[\xi] + T[{\hat\xi}] .
\label{d4}
\end{equation}
The existence of $\hat\xi$ is guaranteed by the invertibility of the triad; we will take
advantage of its nonuniqueness below.   A straightforward computation then gives
\begin{align}
\left\{ {\hat B}[\xi],T[\eta]\right\} &=  
    -\frac{\mu}{2}J[\xi\times\eta] 
    + 2\mu T[\xi\times\eta]   \nonumber\\
    &\quad - \frac{\mu}{2}\int_\Sigma d^2x\,\xi\cdot\eta\,\Delta 
    - T\bigl[\{T[\eta],{\hat\xi}\}\bigr] \approx0\nonumber\\
\left\{ {\hat B}[\xi],B[\eta]\right\} &= 
    2\mu B[\xi\times\eta] -\frac{\mu}{2}J[{\hat\xi}\times\eta]
    + 2\mu T[{\hat\xi}\times\eta] +6\alpha T[\xi\times\eta]  \nonumber\\
    &\quad + \frac{\mu}{2}\int_\Sigma d^2x\,{\hat\xi}\cdot\eta\,\Delta 
    - T\bigl[ \{B[\eta],{\hat\xi}\}\bigr] \approx0\nonumber\\
\left\{{\hat  B}[\xi],\Delta\right\} 
     = &- \epsilon^{abc}\left(\epsilon^{ij}e_{ai}e_{bj}\right)
     \left(\mu{\hat\xi}_c + 9\alpha\xi_c\right)
     -2\mu\epsilon^{abc}\left(\epsilon^{ij}\beta_{ai}e_{bj}\right)\xi_c\nonumber\\
     &- {\hat\xi}^aT_a + \xi^aB_a  - \int_\Sigma d^2x'\, T_a(x')\{\Delta(x),{\hat\xi}^a(x')\} ,
\label{d5}
\end{align}
where $\approx$ means ``weakly equal,'' that is, ``equal up to constraints.''  The first 
two brackets are weakly zero; if $\hat\xi$ can be chosen so that the third is as well, then 
the ${\hat B}[\xi]$ will be first class constraints.

To see that this is possible, we first use the invertibility of $e^a{}_\mu$ to write
\begin{equation}
{\hat\xi}^a = e^{a\mu}\beta_{b\mu}\xi^b + e^{at}\eta ,
\label{d6}
\end{equation}
where $\eta$ is arbitrary.  It is already evident that the last bracket in (\ref{d5}) 
can be made weakly zero, since the right-hand side is linear in $\eta$.  More explicitly,
note that
$$
\epsilon^{\mu\nu\rho}e^a{}_\mu e^b{}_\nu e^c{}_\rho = e\epsilon^{abc} 
$$
with $e = \det|e^a{}_\mu|$, and therefore
$$
\epsilon^{ij}e^a{}_ie^b{}_j = -e\epsilon^{abc}e_c{}^t, \qquad
     \epsilon^{ij}e^b{}_j = -e\epsilon^{abc}e_a{}^ie_c{}^t .
$$
Some simple manipulation then yields
$$
\left\{{\hat  B}[\xi],\Delta\right\} 
     \approx -2\mu eg^{tt}{\eta} + 2\mu e\left(\beta^{tc} - \beta^{ct}\right)\xi_c
     - 2\mu e\left(\beta + \frac{9\alpha}{\mu} \right)e^{ct}\xi_c ,
$$
and we can clearly choose $\eta$ so that the right-hand side vanishes.  More than that, 
it is evident from (\ref{b9}) that $\eta = 0$ on shell.  In that case, ${\hat\xi}^a$ is
identical to the parameter appearing in (\ref{c7}), and ${\hat B}[\xi]$ is essentially the 
generator of diffeomorphisms.

We now consider the remaining constraints, which we can take to be $T^a$ and 
$\Delta$.  It is convenient to write $T^\mu = T^ae_a{}^\mu$.  The Poisson brackets 
(\ref{d2})--(\ref{d3}) then give
\begin{align}
&\left\{ T^i(x),T^j(x')\right\} \approx 
   -\frac{\mu}{2}\epsilon^{ij}\delta^2(x-x') \nonumber\\
&\left\{ T^i(x),T^t(x')\right\} \approx 
   \left\{ T^t(x),T^t(x')\right\} \approx 0 \nonumber\\
&\left\{ T^i(x),\Delta(x')\right\} \approx 
   -\left(\epsilon^{ij}D_j + 2\mu eg^{ti}\right)\delta^2(x-x') \nonumber\\
&\left\{ T^t(x),\Delta(x')\right\} \approx -2\mu eg^{tt}\delta^2(x-x') . \label{d7}
\end{align}
It is clear upon inspection that for any values of $\mu$ and $\alpha$ (except for the
conformal limit $\mu=0$), the matrix of Poisson brackets has a nonzero determinant.
In fact, as I describe in the Appendix, it is not too hard to compute its inverse explicitly.  
Hence no further combination of the $\{T^a,\Delta\}$ gives an additional first class 
constraint.   

We thus have nine canonical pairs of variables ($A^a{}_i$, $e^a{}_i$, and $\beta^a{}_i$), 
six first class constraints ($J^a$ and ${\tilde B}^a$) , and four second class constraints 
($T^a$ and $\Delta$).  Each first class constraint eliminates two phase space degrees
of freedom, while each second class constraint eliminates one \cite{HT}; we therefore have
$18-12-4 = 2$ degrees of freedom left, that is, one canonical pair of free data, describing 
a single local excitation.  While the values of $\mu$ and $\ell$, in the combination $\alpha$, 
affect the algebra of constraints, no choice leads to a change in the types of the constraints 
or a jump in the number of degrees of freedom.  In particular, for the chiral values 
$\mu\ell=\pm1$, these results agree with \cite{GJJ}, while in the asymptotically flat limit 
$\ell\rightarrow\infty$ they go smoothly to the results of \cite{DX}.

\section{Asymptotic symmetries} 

Let us briefly recall a few features of the first-order formulation of (2+1)-dimensional
Einstein gravity with a negative cosmological constant \cite{Wittenb,Banados0,Banados}.
The theory has six first class constraints, which give a canonical representation 
of the underlying symmetries of the theory,  local Lorentz invariance and 
diffeomorphism invariance.  The constraints can be combined to form two mutually 
commuting sets of three generators.  Each set forms a Virasoro algebra, and when 
evaluated at the asymptotic symmetries of anti-de Sitter space, the algebras have
classical central charges.  These central charges, along with the classical conformal 
weights, provide a powerful tool for investigating the boundary conformal field theory.

For topologically massive AdS gravity, we also have two sets of first class 
constraints, $J^a$ and ${\hat B}^a$, which again reflect local Lorentz invariance 
and diffeomorphism invariance.  In general, though, we should not expect these to
split into commuting ``left'' and ``right'' sectors; interactions are likely to
couple the left- and right-movers.  Indeed, from (\ref{d2}),  the symmetry 
generators do not commute: the Poisson brackets of $\hat B$ include a term 
proportional to $J$.   

To understand the central charges and conformal weights, though, it is enough to 
look at a neighborhood of the AdS boundary.  There, from (\ref{b9}) and (\ref{d6}), 
\begin{equation}
{\hat\xi}^a = -\frac{3\alpha}{\mu}\xi^a .
\label{e0}
\end{equation}
If we define
$$
L_\pm[\xi] =  {\hat B}[\xi] + a_\pm J[\xi] ,
$$
it is easy to check that
\begin{align}
\left\{ L_+[\xi],L_-[\eta]\right\}
&= \left\{ {\hat B}[\xi] + a_+J[\xi], {\hat B}[\eta] + a_-J[\eta] \right\}
\nonumber\\
  &= (2\mu - a_+ - a_-) {\hat B}[\xi\times\eta] 
  + (3\alpha -  a_+a_-) J[\xi\times\eta] .
\label{e1}
\end{align}
The right-hand side of (\ref{e1}) will vanish if
$$
a_\pm = \mu\pm\frac{1}{\ell} ,
$$
that is,
\begin{equation}
L_\pm[\xi] = {\hat B}[\xi] 
   + \left(\mu\pm\frac{1}{\ell}\right)  J[\xi] .
\label{e2}
\end{equation}
The remaining Poisson brackets are then
\begin{align}
&\left\{ L_\pm[\xi],L_\pm[\eta]\right\} 
   = \mp {\textstyle\frac{2}{\ell}}L_\pm[\xi\times\eta]  \nonumber\\
&\left\{ L_+[\xi],L_-[\eta]\right\} = 0 . \label{e3a}
\end{align}

An added complication can arise if the parameters  $\xi^a$ are field-dependent:
they may then have nontrivial Poisson brackets with the $L_\pm$, leading to additional 
terms in the algebra (\ref{e3a}).  In particular, we saw earlier that the parameters 
characterizing diffeomorphisms are of the form $\xi^a = e^a{}_\mu\xi^\mu$.   
The algebra thus becomes
\begin{align}
&\left\{ L_\pm[\xi],L_\pm[\eta]\right\} =
    L_\pm\left[ \{L_\pm[\xi],e^a{}_i\}\eta^i 
    - \{L_\pm[\eta],e^a{}_i\}\xi^i \mp {\textstyle \frac{2}{\ell}}(\xi\times\eta)^a\right] 
   \nonumber\\
&\left\{ L_+[\xi],L_-[\eta]\right\} 
   = L_-\left[ \{L_+[\xi],e^a{}_i\}\eta^i\right] - L_+\left[\{L_-[\eta],e^a{}_i\}\xi^i \right] .
\label{e4}
\end{align}
Again, though, matters simplify when we consider only a small neighborhood of 
the AdS boundary.  It is clear that if we could find parameters $\xi$ and $\bar\xi$
such that $\{L_+[\xi],e^a{}_i\} = \{L_-[{\bar\xi}],e^a{}_i\}=0$, the extra terms  
in (\ref{e4}) would vanish.  Globally, this is rarely possible, but we can define asymptotic 
symmetries for which 
\begin{align}
&\left\{ L_+[\xi], e^a{}_i\right\} = - \left(\partial_i\xi^a 
   + \epsilon^{abc}(\omega_{bi} + \frac{1}{\ell}e_{bi})\xi_c\right) \sim 0 \nonumber\\
&\left\{ L_-[{\bar\xi}], e^a{}_i\right\} =  - \left(\partial_i{\bar\xi}^a 
   + \epsilon^{abc}(\omega_{bi} - \frac{1}{\ell}e_{bi}){\bar\xi}_c\right) \sim 0
\label{e6}
\end{align}
at the AdS boundary.\footnote{It is interesting to note that the covariant derivatives 
here are identical to those in the gauge formulation of ordinary Einstein gravity 
\cite{Wittenb,Banados0,Banados}. } We shall see in the next section that these eliminate 
the extra terms in the commutator of $L_+$ and $L_-$ at the boundary.

Equation (\ref{e6}) is easy to solve.  If we choose coordinates such that the leading terms 
in the metric take the form  
$$
ds^2 = \ell^2d\rho^2 + e^{2\rho}\left(\ell^2d\varphi^2 - dt^2\right) ,
$$
we find two families, labeled by functions 
$f(\varphi+t/\ell)$ and ${\bar f}(\varphi-t/\ell)$:
\begin{alignat}{2}
&\xi^0_f = \frac{\ell}{2}e^\rho f   \qquad\qquad
&&{\bar\xi}^0_{\bar f} =  \frac{\ell}{2} e^\rho{\bar f} \nonumber\\
&\xi^1_f = - \frac{\ell}{2}\partial_\varphi f   \qquad\qquad
&&{\bar\xi}^1_{\bar f} =  \frac{\ell}{2} \partial_\varphi {\bar f} \nonumber\\
&\xi^2_f =   \frac{\ell}{2} e^\rho f   \qquad\qquad
&&{\bar\xi}^2_{\bar f}  = - \frac{\ell}{2} e^\rho{\bar f} .
\label{e5}
\end{alignat}
I have chosen a normalization such that the zero-modes of $\xi^t$ and ${\bar\xi}^t$ are
positive and such that
$$
\left[\xi_f,\xi_g\right]^a = \xi_{\{f,g\}}^a, \qquad
\left[{\bar\xi}_{\bar f},{\bar\xi}_{\bar g}\right]^a = -{\bar\xi}_{\{{\bar f},{\bar g}\}}^a, 
$$
where $[\xi,\eta]^\mu = \xi^\nu\partial_\nu\eta^\mu - \eta^\nu\partial_\nu\xi^\mu$ is
the ordinary commutator of (2+1)-dimensional vector fields and 
$\{f,g\} = f\partial_\varphi g - g\partial_\varphi f$ is the commutator of $f$ and $g$ 
viewed as one-dimensional vector fields on the circle.  Not surprisingly, the parameters (\ref{e5})
match those found in ordinary Einstein gravity \cite{Banados0,Banados}, and agree to 
lowest order with the asymptotic AdS Killing  vectors found long ago by Brown and Henneaux 
\cite{BH}.

Restricted to such transformations, the algebra  (\ref{e4}) now becomes
\begin{align}
&\left\{ L_+[\xi_f],L_+[\xi_g]\right\} =  L_+\left[[\xi_f,\xi_g]\right]  + L_+[\chi(f,g)]
   \nonumber\\
&\left\{ L_-[{\bar\xi}_{{\bar f}}], L_-[{\bar\xi}_{{\bar g}}]\right\}
   =  L_-\left[[{\bar \xi}_{\bar f},{\bar \xi}_{\bar g}]\right] 
   + L_-[\chi({\bar f},{\bar g})]
   \nonumber\\
&\left\{ L_+[\xi_f],L_-[{\bar\xi}_{\bar g}]\right\} = 
   -({\hat B} + \mu J)[\chi(f,{\bar g})] - J[{\tilde\chi}(f,{\bar g})] , 
\label{e3b}
\end{align}
with
\begin{alignat}{2}
&\chi^1(f,g) = \frac{\ell}{4}\partial_\varphi\left(f\partial_\varphi g - g\partial_\varphi f\right),
\quad &&\chi^0(f,g) =\chi^2(f,g) = 0  \nonumber\\
&{\tilde\chi}^1(f,g) = \frac{1}{4}\left(f\partial_\varphi^2g + g\partial_\varphi^2f\right) ,
\quad &&{\tilde\chi}^0(f,g) ={\tilde\chi}^2(f,g) = 0 .
\label{ex}
\end{alignat}
We shall see in the next section that the terms involving $\chi$ and $\tilde\chi$ give no
contribution at the AdS boundary.

\section{Boundary terms and central charges}\label{BT}

Up to now, we have focused on the ``bulk'' contributions to the constraints.  We must now  
restore the boundary terms.  Let us first recall a few general features \cite{BH,BHb}.  Consider
a theory of fields $\{\phi_i\}$ in $n+1$ dimensions, with gauge transformations labeled
by parameters $\xi$ and generated by
$$
G[\xi,\phi] = \int_\Sigma d^n\!x\,{\cal G}[\xi,\phi] .
$$ 
Up to boundary terms, these generators should satisfy the appropriate gauge algebra
$$
\left\{ G[\xi,\phi], G[\eta,\phi] \right\} = G[\{\xi,\eta\},\phi] ,
$$
where $\{\xi,\eta\}$ is the Lie bracket for the gauge group.
  
Now let us restore the boundary terms.  Under a general variation of the fields, 
$$
\delta G[\xi,\phi] = \int_\Sigma d^n\!x\, \frac{\delta{\cal G}}{\delta\phi_i}\delta\phi_i
    + \int_{\partial\Sigma}d^{n-1}\!x\, B[\xi,\phi,\delta\phi] .
$$
If the boundary term $B$ is nonzero, $G$ is said to not be ``differentiable.''  In particular, the 
presence of $B$ will lead to delta-function singularities in the Poisson brackets.  It may be
possible to generalize the algebra to include such boundary singularities \cite{Solo,Bering},
but it is normally simpler to choose boundary conditions such that $B$ is itself a total 
variation,
$$
B[\xi,\phi,\delta\phi] = -\delta Q[\xi,\phi]  .
$$
The combination ${\bar G}[\xi,\phi] = G[\xi,\phi]  + Q[\xi,\phi] $ will then have a well-defined
variation, with no boundary terms, and it is easy to show that
\begin{align}
\left\{{\bar G}[\xi,\phi],{\bar G}[\eta,\phi]\right\}
   &=  \iint d^n\!x'\, d^n\!x\,  
  \frac{\delta{\cal G}[\xi,\phi]}{\delta\phi_i(x)}
   \frac{\delta{\cal G}[\eta,\phi]}{\delta\phi_j(x')}\{\phi_i(x),\phi_j(x')\} \nonumber\\
   &= {\bar G}[\{\xi,\eta\},\phi] + K(\xi,\eta) .
\label{k1}
\end{align}

The central term $K(\xi,\eta)$ arises from boundary terms in the integrals, and need not vanish.  It
is most easily evaluated by considering the algebra (\ref{k1}) for the ``vacuum'' configuration, 
for which the boundary charges $Q$ vanish; the right-hand side of (\ref{k1}) then consists solely 
of the central term.

To apply this general formalism to our case, we must first return to (\ref{d1}) and (\ref{d2}) 
and keep track of any boundary terms.  A straightforward calculation yields
\begin{align}
&\left\{J[\xi],J[\eta]\right\} 
    = \dots + \frac{2}{\mu}\int_{\partial\Sigma}\xi^aD_\varphi\eta_a d\varphi \nonumber\\
&\left\{J[\xi],T[\eta]\right\} 
   = \dots - \int_{\partial\Sigma}(\xi\times\eta)_ae^a{}_\varphi d\varphi \nonumber\\
&\left\{J[\xi],B[\eta]\right\} 
   = \dots - \int_{\partial\Sigma}(\xi\times\eta)_a\beta^a{}_\varphi d\varphi \nonumber\\
&\left\{T[\xi],T[\eta]\right\} = \dots \nonumber\\
&\left\{B[\xi],T[\eta]\right\} 
   = \dots + \int_{\partial\Sigma}\left[ \xi^aD_\varphi\eta_a 
   + 2\mu(\xi\times\eta)_ae^a{}_\varphi\right] d\varphi \nonumber\\
&\left\{B[\xi],B[\eta]\right\} 
   = \dots + \int_{\partial\Sigma}(\xi\times\eta)_a\left(2\mu\beta^a{}_\varphi 
   + 6\alpha e^a{}_\varphi\right)d\varphi ,
\label{k2}
\end{align}
where the omitted bulk terms are all proportional to the constraints, and vanish weakly. 
For asymptotically anti-de Sitter boundary conditions, we see from (\ref{b9}) that   
$\beta^a = -\frac{3\alpha}{\mu}e^a$ and ${\hat\xi}^a = -\frac{3\alpha}{\mu}\xi^a$
at the boundary.  Some simple algebra then gives
\begin{alignat}{2}
\left\{ L_\pm[\xi],L_\pm[\eta]\right\}& &&= \dots 
   \pm\frac{4}{\ell} \int_{\partial\Sigma}\left[ 
  \left(1 \pm \frac{1}{\mu\ell}\right)\xi^a D_\varphi \eta_a
  + \frac{3\alpha}{\mu}(\xi\times\eta)^a e_{a\varphi}\right]
  d\varphi \nonumber\\
&&&= \dots \pm\frac{4}{\ell}\left( 1 \pm \frac{1}{\mu\ell}\right)
    \int_{\partial\Sigma}\xi^a\left[ \partial_\varphi\eta_a
    + \epsilon_{abc}\left(\omega^b{}_\varphi \pm \frac{1}{\ell}e^b{}_\varphi\right) \eta^c
    \right] d\varphi \nonumber\\
\left\{ L_+[\xi],L_-[\eta]\right\}& &&= \dots .
\label{f3}
\end{alignat}
Evaluated at the AdS ``vacuum'' state, the right-hand sides of these
expressions are the central terms $K_\pm$.

If our asymptotic symmetries (\ref{e5}) were exact---that is, if (\ref{e6}) were 
satisfied exactly, and not just asymptotically---then the integrands on the right-hand 
side of (\ref{f3}) would vanish.  But the symmetries are not quite exact, and a simple 
calculation shows that
\begin{align}
&\left\{ L_+[\xi_f],L_+[\xi_g]\right\} =  \dots
   + \frac{\ell}{32\pi G}\left( 1 + \frac{1}{\mu\ell}\right)
    \int_{\partial\Sigma}\left( \partial_\varphi f \partial^2_\varphi g 
    - \partial_\varphi g \partial^2_\varphi f \right) d\varphi \nonumber\\
   \nonumber\\
&\left\{ L_-[{\bar\xi}_{{\bar f}}], L_-[{\bar\xi}_{{\bar g}}]\right\} = \dots
   - \frac{\ell}{32\pi G}\left( 1 - \frac{1}{\mu\ell}\right)
   \int_{\partial\Sigma}\left( \partial_\varphi {\bar f} \partial^2_\varphi {\bar g} 
    - \partial_\varphi {\bar g} \partial^2_\varphi {\bar f} \right) d\varphi ,
\label{fx}
\end{align}
where I have restored the factors of $16\pi G$.  These are precisely the central terms
for two Virasoro algebras with central charges
$$
c_\pm = \frac{3\ell}{2G}\left(1\pm\frac{1}{\mu\ell}\right),
$$
matching the results (\ref{intro1}) that had been previously obtained using very 
different methods \cite{KL,Solodukhin,Hotta}.

Finally, let us directly evaluate the boundary terms $Q_{L_\pm}$.  Here we can use
some results from pure Einstein gravity, where the same problem was discussed
in \cite{Banados0,Banados}.  Note first that from (\ref{c3}) and (\ref{e2}), the
boundary terms in the variation of $L_\pm$ are
\begin{align}
\delta L_\pm[\xi] 
&= \dots - \int_{\partial\Sigma} \left[
    \xi^a\delta\beta_{a\varphi} + {\hat\xi}^a\delta e_{a\varphi}
   + \frac{2}{\mu}\left(\mu\pm\frac{1}{\ell}\right)\xi^a\delta A_{a\varphi}
   \right]d\varphi \nonumber\\
&= \dots - \int_{\partial\Sigma} \xi^\mu\left[
   e^a{}_\mu\delta\beta_{a\varphi} + \beta^a{}_\mu\delta e_{a\varphi}
   + \frac{2}{\mu}\left(\mu\pm\frac{1}{\ell}\right)e^a{}_\mu\delta A_{a\varphi}
   \right]d\varphi .
\label{l1}
\end{align}
As before, anti-de Sitter boundary conditions require that
$\beta^a = -\frac{3\alpha}{\mu}e^a$, and a bit of algebra reduces (\ref{l1}) to
\begin{align}
\delta L_\pm[\xi] 
&= \dots - \int_{\partial\Sigma} \xi^\mu\left[
    -\frac{6\alpha}{\mu}e^a{}_\mu\delta e_{a\varphi} 
    + 2 \left( 1 \pm \frac{1}{\mu\ell}\right)e^a{}_\mu\delta A_{a\varphi} 
     \right]d\varphi  \nonumber\\
&= \dots - 2\left( 1 \pm \frac{1}{\mu\ell}\right) \int_{\partial\Sigma}
    \xi^\mu e^a{}_\mu \,\delta\left(
    \omega_{a\varphi}\pm\frac{1}{\ell}e_{a\varphi}\right)d\varphi .
\label{l2}
\end{align}
We now adopt the boundary conditions of \cite{Banados0,Banados}, which translate
to
\begin{equation*}
\omega^a{}_t =  \frac{1}{\ell^2}e^a{}_\varphi, \qquad
\omega^a{}_\varphi =  e^a{}_t, \qquad \delta e^a{}_\rho = 0 ,
\label{l3}
\end{equation*} 
and note that for our asymptotic symmetries, $\xi^\varphi = \pm\frac{1}{\ell}\xi^t$.  
The variation (\ref{l2}) is thus
\begin{align*}
\delta L_\pm[\xi] 
&= \dots - 2\left( 1 \pm \frac{1}{\mu\ell}\right) \int_{\partial\Sigma}
    \left[\xi^t\left(e^a{}_t\pm\frac{1}{\ell}e^a{}_\varphi\right) + \xi^\rho e^a{}_\rho\right]
    \delta \left(e_{at}\pm \frac{1}{\ell}e_{a\varphi}\right) d\varphi \nonumber\\
&= \dots -\left( 1 \pm \frac{1}{\mu\ell}\right) \delta \int_{\partial\Sigma}
    \left[\xi^t \left(e^a{}_t\pm \frac{1}{\ell}e^a{}_\varphi \right)
    + 2 \xi^\rho\, e^a{}_\rho\right]
    \left(e_{at}\pm \frac{1}{\ell}e_{a\varphi}\right) d\varphi ,
\label{l4}
\end{align*}
and the $L_\pm$ are thus differentiable if we add boundary terms
\begin{equation}
Q_\pm[\xi] =  \frac{1}{16\pi G}\left( 1 \pm \frac{1}{\mu\ell}\right) \int_{\partial\Sigma}
    \left[\xi^t \left(e^a{}_t\pm \frac{1}{\ell}e^a{}_\varphi \right)
    + 2 \xi^\rho\, e^a{}_\rho\right]
    \left(e_{at}\pm \frac{1}{\ell}e_{a\varphi}\right) d\varphi .
\label{l5}
\end{equation}

These boundary terms are identical to those of ordinary Einstein gravity, except
for the prefactors of $1 \pm \frac{1}{\mu\ell}$.   That is,
\begin{equation}
Q_\pm^{\hbox{\tiny TMG}}[\xi] 
    = \left( 1 \pm \frac{1}{\mu\ell}\right) Q_\pm^{\hbox{\tiny Einstein}}[\xi] ,
\label{l6}
\end{equation}
in agreement with \cite{KL,Solodukhin}.  Further, we can now verify the claim in 
the preceding section that the $\chi$ and $\tilde\chi$ terms in (\ref{ex})
are irrelevant at the boundary.  Indeed, these terms only appear in (\ref{l5}) in
the form $\chi^\rho(g_{\rho t}\pm \frac{1}{\ell}g_{\rho\varphi})$, and
vanish by virtue of our boundary conditions.

\section{Chirality}

It has recently been argued that topologically massive AdS gravity is chiral at 
the critical coupling $\mu\ell=\pm1$ \cite{Stromingerb}.  In the present context, 
this feature can be understood as follows.  

Consider first a generic coupling, and let $\xi^\mu$ be a vector field that satisfies 
the fall-off conditions (\ref{e5}) but is nonzero at the boundary.  From (\ref{fx}), 
the constraints $L_\pm[\xi]$ are no longer first class:  their Poisson brackets 
are not weakly zero.  Constraints that are not first class do not generate gauge
transformations, but rather determine asymptotic symmetries \cite{Benguria}.  
Hence some configurations that are formally diffeomorphic will nevertheless be
physically inequivalent---they will differ by a symmetry rather than a gauge 
equivalence.   As a consequence,  new ``would-be pure gauge'' degrees of freedom 
appear at the boundary, which are conjecturally the source of the degrees of 
freedom of the black hole \cite{Carlip,Carlipc,Carlipd}.

If $\mu\ell=1$, on the other hand---or, by an obvious extension, $\mu\ell=-1$---it 
is apparent from (\ref{fx}) and (\ref{l5}) that $c_-$ and $Q_-$ vanish.  Thus 
$L_-[\xi]$ remains first class even at the boundary, and one chirality of diffeomorphisms 
extends to the boundary as a true gauge invariance.  This eliminates one chiral sector 
of the ``massless gravitons'' discussed in \cite{LSS}.  The remaining asymptotic 
symmetry group consists of only one copy of the Virasoro algebra, and the boundary 
theory is thus chiral.

Note, however, that this argument does not eliminate bulk excitations that are not
diffeomorphic to zero in the interior.  In particular, the linearized excitations of
\cite{CDWW} and \cite{GKP} yield solutions with nonconstant curvature.  No 
diffeomorphism, whether or not it extends to the boundary, can remove such 
excitations.

\section{Conclusions}

This work has, first of all, established the existence of a local degree of  freedom 
in topologically massive AdS gravity at all values of the couplings.  In 
particular, I confirm the results of  \cite{GJJ} for the chiral coupling $\mu\ell=\pm1$.  
The constraint analysis presented here is, in a sense, complementary to the 
perturbative analysis of \cite{CDWW,GKP}.  Those papers show that weak 
field solutions exist and remain well-behaved at the AdS boundary, but cannot 
address effects beyond the weak field approximation, while the present analysis 
is fully nonperturbative, but does not address boundary behavior.  Since the weak
field perturbations have negative energy (relative to the black hole), these results
together provide a strong indication that the theory is unstable.

On the other hand, this work also confirms that the boundary central charges of
topologically massive AdS gravity are shifted, and that at the chiral coupling,
one of the two central charges vanishes.  This presents a bit of a puzzle for the
AdS/CFT correspondence: the central charge measures the number of states in
the dual conformal field theory, and the vanishing of a central charge should
mean, in some sense, that some fields disappear.   

Note, though, that the \emph{total} central charge, $c_++c_-$, is independent
of $\mu$; the vanishing of $c_-$ at $\mu\ell=1$ is compensated by an increase
in $c_+$.  The same behavior can be seen in the boundary conformal weights:  
by (\ref{l5}), when $c_-$ and $Q_-$ vanish, $Q_+$ doubles.  For the BTZ black 
hole, this is reflected in the fact that all solutions are extremal at the chiral 
coupling \cite{Moussa,LSS}, while the entropy nevertheless remains independent 
of $\mu$.  How this feature is manifested in the bulk---where the constraint algebra, 
at least, shows no special behavior as couplings vary---remains a mystery.

The value of the total central charge also presents a second puzzle: it is the 
same for topologically massive gravity as it is for ordinary Einstein gravity.  
The counting of states via the Cardy formula will thus match the results of 
Einstein gravity, which are already sufficient to account for for the BTZ black
hole entropy; we will see no additional contribution from the ``massive 
graviton'' at \emph{any} value of the coupling constant.  This should not 
really be such a surprise, though: the classical contribution (\ref{intro1}) to 
the central charge is really of order ${\cal O}(1/\hbar)$, while an ordinary 
propagating field contributes ${\cal O}(1)$.  This suggests that the classical 
Poisson bracket analysis of the boundary conformal field theory might not 
capture enough information to tell us about the massive graviton degrees
of freedom, which may only appear at higher orders in $\hbar$.

Can chiral topologically massive gravity be saved?  The negative-energy weak 
field excitations of \cite{CDWW} can be built from compactly supported initial 
data---that is, they represent arbitrarily small and arbitrarily localized 
perturbations, which cannot be excluded by boundary conditions in any obvious 
way.  The constraint analysis developed here further shows that these perturbations
represent the ``right amount'' of initial data, one free phase space degree of 
freedom per point.  It remains conceivable, however, that higher order corrections
to the weak field solutions violate Fefferman-Graham boundary conditions, or
lead to a finite lower bound to the negative energies that appear perturbatively.
Unfortunately, the one known positive energy theorem for topologically massive 
AdS gravity, which follows from the existence of a supersymmetric extension,   
goes in the wrong direction \cite{Deser,Abbott}: with the sign choice for
which the black hole has positive mass, the energy of local excitations is strictly 
nonpositive.  Nevertheless, a more detailed investigation of boundary conditions
beyond first order perturbation theory could be of interest.

\vspace{1.5ex}
\begin{flushleft}
\large\bf Acknowledgments
\end{flushleft}
\noindent 
I would like to thank Stanley Deser, Marc Henneaux, Andrew Waldron, and Derek 
Wise for helpful discussions.
This work was supported in part by the Department of Energy under grant
DE-FG02-91ER40674.

\appendix
\section{}

\noindent This Appendix contains additional details of some calculations.\\[-.7ex]

\noindent{\bf Classical equations of motion:}\\[-.7ex]

The first-order equations of motion (\ref{b7}) in the variables $\{A,e,\beta\}$ are 
equivalent to the standard field equations for topologically massive gravity.  To
see this, note initially that the second equation in (\ref{b7}) is simply the torsion 
constraint (\ref{b2}), with the spin connection $\omega$ defined in terms of $A$ 
and $e$ by (\ref{b5}).  This constraint determines the spin connection as a function 
of the triad.  The first equation in (\ref{b7}) then becomes, in component form,
\begin{align*}
&\epsilon^{abc}F_{a\mu\nu} = 
    -\frac{\mu}{4}\left( \beta^b{}_\mu e^c{}_\nu - \beta^b{}_\nu e^c{}_\mu
    - \beta^c{}_\mu e^b{}_\nu + \beta^c{}_\nu e^b{}_\mu\right) \\
&F_{a\mu\nu} = R_{a\mu\nu} + \frac{\mu^2}{2}\epsilon_{abc}e^b{}_\mu e^c{}_\nu
   \quad\hbox{with}\  
   R_a = d\omega_a + \frac{1}{2}\epsilon_{abc}\omega^b\wedge\omega^c .
\end{align*}
Contracting with $e_c{}^\nu$, and noting that $e_c{}^\nu\epsilon^{abc}R_{a\mu\nu} 
= \frac{1}{2}R^b{}_\mu$ (where  $R^b{}_\mu$ is the Ricci tensor), we find that
$$
\beta^a{}_\mu =
  -\frac{2}{\mu}\left(R^a{}_\mu - \frac{1}{4}e^a{}_\mu R 
  + \frac{\mu^2}{2}e^a{}_\mu\right) .
$$
Upon inserting this expression into the last equation in (\ref{b7}), a little algebra yields
\begin{equation}
G^\mu{}_\sigma - \frac{1}{\ell^2}\delta^\mu_\sigma 
   - \frac{1}{\mu}\epsilon^{\mu\nu\rho}\nabla_\nu\left(
   R_{\rho\sigma} - \frac{1}{4}g_{\rho\sigma}R\right) = 0 ,
\label{b8}
\end{equation}
the usual field equations for topologically massive AdS gravity.  Contraction gives
$R=-{6}/{\ell^2}$, which implies in turn that
\begin{align*}
&\beta^a{}_\mu =
  -\frac{2}{\mu}\left(R^a{}_\mu + \frac{2}{\ell^2}e^a{}_\mu 
  + \frac{3}{2}\alpha e^a{}_\mu\right) 
   \nonumber\\
&\beta = \beta^a{}_\mu e_a{}^\mu = -\frac{9}{\mu}\alpha ,
\end{align*}
which we can recognize as equation (\ref{b9}).\\

\noindent{\bf Poisson brackets of the constraints:}\\[-.7ex]

The Poisson brackets of the smeared constraints $\{J[\xi],T[\xi],B[\xi]\}$ with
with the canonical variables $\{A,e,\beta\}$ are easily computed from the 
fundamental brackets (\ref{c1}).  One finds
\begin{align}
\left\{ J[\xi],A^a{}_i\right\} &= -D_i\xi^a \nonumber\\
\left\{ J[\xi],e^a{}_i\right\} &= -\epsilon^a{}_{bc}e^b{}_i\xi^c \nonumber\\
\left\{ J[\xi],\beta^a{}_i\right\} &= -\epsilon^a{}_{bc}\beta^b{}_i\xi^c \nonumber\\
\left\{ T[\xi],A^a{}_i\right\} &= -\frac{\mu}{2}\epsilon^a{}_{bc}e^b{}_i\xi^c \nonumber\\
\left\{ T[\xi],e^a{}_i\right\} &= 0 \nonumber\\
\left\{ T[\xi],\beta^a{}_i\right\} &= -D_i\xi^a + 2\mu\epsilon^a{}_{bc}e^b{}_i\xi^c \nonumber\\
\left\{ B[\xi],A^a{}_i\right\} &= -\frac{\mu}{2}\epsilon^a{}_{bc}\beta^b{}_i\xi^c\nonumber\\
\left\{ B[\xi],e^a{}_i\right\} &=  -D_i\xi^a + 2\mu\epsilon^a{}_{bc}e^b{}_i\xi^c \nonumber\\
\left\{ B[\xi],\beta^a{}_i\right\} &= 2\mu\epsilon^a{}_{bc}\beta^b{}_i\xi^c
    + 6\alpha\epsilon^a{}_{bc}e^b{}_i\xi^c , \label{c5}
\end{align}
where $D$ is the gauge-covariant exterior derivative for the connection $A$.\\

\noindent{\bf Inverting the second class constraints}\\[-.7ex]

In section \ref{algebra}, it was shown that the constraints $\{T^a,\Delta\}$ were
second class, that is, that the matrix of their Poisson brackets was nonsingular.
Here I compute the inverse of that matrix explicitly.  Let us write $T^A = (T^a,\Delta)$,
and define
$$
K^{AB}(x,x') = \left\{T^A(x),T^B(x')\right\} .
$$
From equations (\ref{d2}) and (\ref{d3}), we have
\begin{align}
&K^{ab}(x,x') = -\epsilon^{abc}u_c\delta^2(x-x') \nonumber\\
&K^{a\Delta}(x,x') = -\epsilon^{jk}e^a{}_j(x)\partial_k\delta^2(x-x')
   + 4u^a\delta^2(x-x') \nonumber\\
&K^{\Delta a}(x,x') = \epsilon^{jk}\partial_j\left(e^a{}_k\delta^2(x-x')\right)
   - 4u^a\delta^2(x-x')  \nonumber\\
&K^{\Delta\Delta}(x,x') = 0 ,
\label{h1}
\end{align}
where I define
$$
u_a = -\frac{\mu}{4}\epsilon_{abc}\epsilon^{ij}e^b{}_ie^c{}_j , \qquad
P^a_b = \delta^a_b - \frac{u^au_b}{u^2} .
$$
We wish to find the inverse kernel $K^{-1}_{AB}$, which should satisfy
\begin{align}
&\int d^2u\left[ K^{-1}_{ab}(x,u)K^{bc}(u,x') 
   + K^{-1}_{a\Delta}(x,u)K^{\Delta c}(u,x')\right] = \delta^c_a\delta^2(x-x')
   \nonumber\\
&\int d^2u\left[ K^{-1}_{ab}(x,u)K^{b\Delta}(u,x') \right] =  
  \int d^2u\left[ K^{-1}_{\Delta b}(x,u)K^{bc}(u,x') \right] = 0 \nonumber\\
&\int d^2u\left[ K^{-1}_{\Delta b}(x,u)K^{b\Delta}(u,x') \right] = \delta^2(x-x')
\label{h2}
\end{align}
A tedious but straightforward calculation gives
\begin{align}
K^{-1}_{\Delta b}(x,x') &= -K^{-1}_{b\Delta}(x,x') 
   = \frac{1}{4}\frac{u_b}{u^2}\delta^2(x-x') \nonumber\\
K^{-1}_{ab}(x,x') &= -\epsilon_{abc}\frac{u^c}{u^2}\delta^2(x-x') \nonumber\\
  &\quad + \frac{1}{2\mu}\left[ \frac{u_a}{u^2}(x)\left(P_b{}^ce_c{}^j\right)(x')
  + \frac{u_b}{u^2}(x')\left(P_a{}^ce_c{}^j\right)(x)\right]\partial_j\delta^2(x-x') .
\label{h3}
\end{align}
Checking (\ref{h2}) is now fairly easy, if one notes that
$$
u_ae^a{}_i=0 , \qquad 
\epsilon^{jk}e^a{}_k = \frac{2}{\mu}\epsilon^{abc}e_b{}^ju_c  .
$$
The inverse kernel $K^{-1}_{AB}$ is clearly nonsingular unless $u^2=0$.  But it
is easy to check that $u^2\sim\det|g_{ij}|$, so this can only occur for singular
metrics.\\

\end{document}